\documentclass[11pt]{article}
\usepackage[latin1]{inputenc}

\usepackage[top=1.1in, bottom=1.1in, left=1.15in, right=1.15in]{geometry}
\usepackage{setspace}
\usepackage[pdftex]{graphics}
\usepackage{graphicx}
\usepackage{psfrag}
\usepackage{color}
\usepackage{amsfonts}
\usepackage{amssymb, amsmath}
\usepackage{amsthm}
\usepackage{subcaption}
\usepackage{float}
\usepackage{bm}
\DeclareGraphicsExtensions{.pdf,.png,.jpg}

\usepackage{natbib} 
\usepackage{enumerate}

\usepackage[colorlinks=true, pdfstartview=FitV, linkcolor=black, citecolor=black, urlcolor=black]{hyperref}

\def\E{\textnormal{E}}

\newtheorem{theorem}{Theorem}
\newtheorem{proposition}{Proposition}
\newtheorem{lemma}{Lemma}

\newtheorem{assumption}{Assumption}

\newtheorem{remark}[theorem]{Remark}

\begin{document}
	
\title{A first-stage representation for instrumental variables \\ quantile regression}

	\author{Javier Alejo\footnote{IECON-Universidad de la Rep\'ublica, Montevideo, Uruguay. E-mail: javier.alejo@ccee.edu.uy} 
\and 
Antonio F. Galvao\footnote{Michigan State University, East Lansing, USA. E-mail: agalvao@msu.edu}
\and
Gabriel Montes-Rojas\footnote{CONICET and Instituto Interdisciplinario de Econom\'ia Pol\'itica, Universidad de Buenos Aires, Ciudad Aut\'onoma de Buenos Aires, Argentina. E-mail: gabriel.montes@fce.uba.ar}	}
	
\maketitle
	
\begin{abstract}
This paper develops a first-stage linear regression representation for the instrumental variables (IV) quantile regression (QR) model. The quantile first-stage is analogous to the least squares case, i.e., a linear projection of the endogenous variables on the instruments and other exogenous covariates, with the difference that  the QR case is a weighted projection. The weights are given by the conditional density function of the innovation term in the QR structural model, conditional on the endogeneous and exogenous covariates, and the instruments as well, at a given quantile. We also show that the required Jacobian identification conditions for IVQR models are embedded in the quantile first-stage. We then suggest inference procedures to evaluate the adequacy of instruments by evaluating their statistical significance using the first-stage result. The test is developed in an over-identification context, since consistent estimation of the weights for implementation of the first-stage requires at least one valid instrument to be available. Monte Carlo experiments provide numerical evidence that the proposed tests work as expected in terms of empirical size and power in finite samples. An empirical application illustrates that checking for the statistical significance of the instruments at different quantiles is important. The proposed procedures may be specially useful in QR since the instruments may be relevant at some quantiles but not at others.

\vspace{7mm}
		
\vspace{3mm}
		
\end{abstract}
	
\doublespacing 
\newpage

\section{Introduction}

Instrumental variables (IV) methods are one of the main workhorses to estimate causal relationships in empirical analysis. Standard IV regression methods stress that for instruments to be valid they must be exogenous and must be related to the endogenous variables. The latter condition is usually evaluated by using a first-stage auxiliary regression, where a linear model is used to make inference on the degree of association of the IV and the endogenous variables. While this is usually accepted as a valid procedure, its representation is in fact specific to the two-stage least squares (2SLS) model for mean models. This paper derives a first-stage representation for quantile regression (QR) models.

Several IV methods have been proposed in QR to solve endogeneity when the covariates are correlated with the error term in a regression model. \citet{CH04, CH05,CH06,CH08} (CH hereafter) develop an instrumental variables quantile regression (IVQR) procedure that has been applied in several contexts. It is one of the most prolific approaches in terms of subsequent work, as it provides a general procedure to use IV for endogeneity of regressors \citep[see, e.g.,][]{CHJ07, CHJ09, Galvao11, Chetverikov16}. We refer to \citet{CH20} for an overview of IVQR.\footnote{There is also a more recent literature on GMM QR, see e.g., \citet{Firpoetal21} and references therein.}
   
CH comment that their method is a simple solution to a 2SLS analog.\footnote{This has been formally established in \citet{GalvaoMontes15}.} However, the first-stage of the IVQR estimator has not been explicitly considered, as it is implemented as an inverse QR estimator. The IVQR estimator contrasts to alternative procedures where the first-stage is implemented. For instance, \citet{Amemiya82}, \citet{Powell83}, \citet{ChenPortnoy96}, and \citet{KimMuller04} use an explicit first-stage that fits the endogenous variable(s) as a function of exogenous covariates and IV, and this is then plugged in a second-stage. \citet{Lee07} also adopts a two-step control-function approach where in first step consists of estimation of the residuals of the reduced-form equation for the endogenous explanatory variable. \citet{MaKoenker06} present an estimator for a recursive structural equation model. 

This paper builds on the IVQR estimator and shows that a first-stage regression model can be explicitly recovered from the CH IVQR estimator. The first-stage IVQR (FS-IVQR) is a linear projection of the endogenous variables on the instruments and other exogenous variables, with the difference that the QR case is a weighted regression, that is, it has the representation of a weighted least squares (WLS) regression of the endogenous variable(s) on the IV and the exogenous regressors. The weights are given by the conditional density function of the innovation term in the QR structural model, conditional on the endogeneous and exogenous covariates together with the instruments, at a given quantile. This result provides a clear analogy between the first-stage in 2SLS and IVQR. The derivation of the result is simple. We write the IVQR model as a constrained Lagrangian optimization problem and show that one of the restrictions that must be satisfied is the analogue of the first-stage. 

The CH IVQR method requires an identification condition that is based on the full-rank of the Jacobian for the exogeneity of the instruments.  The lack of identification when the Jacobian is not full rank implies that estimating the parameters can be extremely difficult and the first-order asymptotics can be a poor guide of the actual sampling distributions \citep[see e.g.][]{Dufour97}. In this paper, we show that a necessary condition for the Jacobian identification conditions for IVQR models are embedded in the quantile first-stage representation. Hence, the FS-IVQR representation is directly related to the Jacobian requirement of CH IVQR.

We propose a two-step FS-IVQR estimator. The practical implementation of the estimator is straightforward and as follows. First, from the IVQR one estimates the conditional density function at a selected quantile, which produces an estimate of the weights. The weighting factor is estimated from the IVQR errors using, for instance, sparsity or kernel methods (see, e.g., \cite{Koenker05}). Second, a standard WLS regression is implemented by regressing the endogenous variable on the instruments and exogenous variables with weights from the first step -- this is parallel to the first-stage model used in 2SLS, but using weights. We derive the limiting distribution of the two-step FS-IVQR estimator and show that, under some standard regularity conditions, it is asymptotically normal.

The first-stage regression for conditional average models has been used as a natural framework to evaluate the validity of instruments since one can test for their statistical significance, that is, how the IV impact the endogenous variable(s). Based on the proposed FS-IVQR model, we suggest an analogous test procedure to assess the validity of the IV for given quantiles. In particular, we test the statistical significance of the FS-IVQR coefficients. This is a test for the Jacobian identification condition with the null hypothesis that the required rank condition is not satisfied -- the coefficients are equal to zero. A simple Wald statistic can be used to test this null, and we show that it has an asymptotic Chi-square distribution. Nevertheless, the implementation of the test requires consistent estimation of the weights, which in turn requires at least one valid available instrument. The requirement of consistent estimation of weights in the first-stage QR exposes a caveat with the IVQR model, and hence, we suggest practical use of the test in an over-identification context, that is, when at least one valid instrument is available. In spite of this issue, testing using the first-stage QR allows for a procedure in empirical work to evaluate the degree of association of the IV to the endogenous variable that is parallel to the standard first-stage in two-stage least squares (FS-2SLS).

The requirement of consistent estimation of weights in the QR first-stage leads to two important conclusions of this paper. First, it is difficult to derive an analogous F-statistic type rule-of-thumb for categorizing weak instruments as in, among others, \citet{Staiger97}, \citet{Sanderson16}, and \citet{Lee20}, for ordinary least-squares (OLS) models (see \citet{StockYogo05} for an extensive discussion). A complete parallel testing procedure to evaluate the validity of IV in the QR case only works when one is able to estimate the structural parameters -- and consequently the weights -- consistently, which in turn requires at least one valid instrument. Therefore, dropping the valid instrument requirement is a very interesting line of future research for investigating the weak instruments problem. Second, an important conclusion for practical work is that weak identification robust inference procedures, as in \citet{CH08}, \citet{Jun08}, \citet{CHJ09}, and \cite{AndrewsMikusheva16}, are a very important avenue for empirical applications using QR instrumental variables models. We strongly suggest the use of the QR first-stage and testing proposed here along with the weak identification robust inference procedures in empirical applications.

One important feature of the procedure developed in this paper is that instruments could be statistically insignificant in FS-2SLS, but they could still be related to the endogenous variable in the IVQR set-up. The reason is that the FS-2SLS test only evaluates a mean effect, but the FS-IVQR, because of its specific weighting procedure, allows for different first-stage coefficients across quantiles.  As a result, the IV could be relevant at some quantiles but not for the mean (and vice-versa), an issue that has been discussed in \citet{Chesher03} and subsequent literature.  The test developed here thus allows inference on the validity of the IV for the exogeneity condition across quantiles, rather than only a mean effect.

We use a Monte Carlo exercise to evaluate the finite sample performance of the proposed tests. The tests have correct size in all cases studied, where the structural parameters can be consistently estimated under the null hypothesis. We consider alternative cases where there is no identification under the null. The tests have excellent power properties. In particular these experiments highlight the case where the FS-2SLS test for the mean-based model suggests the instrument is not valid, but the proposed FS-IVQR procedure finds it is for some quantiles.

As an empirical illustration, we apply the FS-IVQR estimator to the \citet{Card95} data on instrumenting education using college proximity. The analysis reveals heterogeneity in the significance of the IV across quantiles. In fact, while the 2SLS analysis shows that one instrument (proximity to 2-year college) is not statistically significant in the first-stage, it is indeed for high quantiles.

The paper is organized as follows. Section \ref{constrainedCH} briefly reviews the CH IVQR estimator, rewrites that estimator as a constrained minimization problem and derives the first-stage representation for the IVQR. Then it shows that the FS-IVQR estimator is equivalent to the identification condition in CH. Section \ref{test} presents the first-stage test for validity of instruments.
Section \ref{estimator.and.asymptotics} discusses its empirical implementation and derives the estimators' asymptotic distribution.
 Section \ref{MonteCarlo} provides finite sample Monte Carlo evidence. Section \ref{empirical} applies the proposed tests to an empirical problem. Finally, Section \ref{conclusion} concludes.

\section{A first-stage representation for IVQR}\label{constrainedCH}

\subsection{The IVQR estimator and its variants}

Let $(y,d,x,z)$ be random variables, where $y$ is a scalar outcome of interest, 
$d$ is a $1\times r$ vector of endogenous variables, 
$x$ is a $1\times k$ vector of exogenous control variables, and $z$ is a $1\times p$ vector of exogenous instrumental variables, with $p\geq r$.  Define $w=(x,z)$ and $s=(d,x,z)$. 

\citet{CH06} developed estimation and inference for a generalization of the QR model with endogenous regressors. A linear representation of the model takes the following form
\begin{equation}\label{eq:structural}
y = d\alpha_0(u_{d})+x\beta_0(u_{d}),  \ u_{d}|x,z \sim \textnormal{Uniform}(0,1),
\end{equation}
where $u_{d}$ is the nonseparable error or rank and the subscript indicates the endogenous covariates of the model.
Under some regularity conditions, CH establish the following IV identification function
\begin{equation}\label{eq:Qrestriction}
P[y\leq d\alpha_0(\tau)+x\beta_0(\tau)|x,z]=P[u_{d} \leq \tau | x, z]=\tau.
\end{equation}
Although each parameter and estimator is indexed by the quantile $\tau \in (0,1)$, throughout the paper we will suppress the dependence on $\tau$.

The restriction in \eqref{eq:Qrestriction} can be used to estimate the parameters of interest. For a given quantile $\tau$, the population IVQR estimator for model in \eqref{eq:structural}, is given by
\begin{equation*}
\underset{\alpha}{{\rm min}}{\|\gamma(\alpha)\|_{A}},
\end{equation*}
where
\begin{equation*}
(\beta(\alpha),\gamma(\alpha))=\underset{\beta,\gamma}{\rm argmin}\; \E \left[\rho_\tau(y-d\alpha-x\beta-z\gamma) \right],
\end{equation*}
and $\rho_\tau(u)=u(\tau-\bm 1(u<0))$ is the check function, and $\|\cdot\|_A=\cdot'A\cdot$ is the Euclidean distance for any positively definite matrix $A$ of dimension $p\times p$.  


As noted by \citet[][p.501]{CH06}, the IVQR estimator is asymptotically equivalent to a particular GMM estimator where the QR first order conditions are used  as moment conditions. In particular, it would involve a Z-estimator solving
\begin{align}
\E\left[ x'\left(\bm{1}[y-d\alpha-x\beta<0]-\tau\right)\right] & =\bm{0}_k,\label{m1}\\
\E\left[z'\left(\bm{1}[y-d\alpha-x\beta<0]-\tau\right)\right] &=\bm{0}_p,\label{m2},
\end{align}
where $\bm{1}(\cdot)$ is the indicator function. Here $\bm{0}_k$ and $\bm{0}_p$ are null vectors with dimensions $k\times 1$ and $p\times 1$, respectively.

Different estimators have been proposed in the GMM framework based on identifying the structural parameters from equations \eqref{m1}--\eqref{m2}. \citet{Kaplan17}, \citet{ChenLee18} and  \citet{dCGKL19} provide general estimation procedures based on smoothing techniques of the non-differentiable indicator function. However, the constructed estimator differs from the CH IVQR one. This can be seen in the fact that the term $z\gamma$ is not considered altogether from the regression model. Our procedure follows the CH estimator and their specific notation.

\subsection{The IVQR estimator as a constrained minimization problem}

The IVQR estimator proposed by \citet{CH06}, for a given quantile $\tau$, can be written as a constrained minimization problem, where the constraints are the moment conditions, that is,
\begin{equation}
\underset{(\alpha,\beta,\gamma)}{{\rm min}}{\|\gamma\|_A},
\end{equation}
subject to
\begin{align}
\E\left[x'\left(\bm{1}[y-d\alpha-x\beta-z\gamma<0]-\tau\right)\right] & =\bm{0}_k,\\
\E\left[z'\left(\bm{1}[y-d\alpha-x\beta-z\gamma<0]-\tau\right)\right] &=\bm{0}_p.
\end{align}

Now we write this constrained optimization as a Lagrangian problem\footnote{See \cite{Pouliot19} and \cite{Kaido21} for recent contributions that tackle the problem of practical implementation of the IVQR methods.} as
\begin{align}
\mathcal{L}(\alpha,\beta,\gamma,\lambda_x,\lambda_z)&=\|\gamma\|_A
+\lambda_x \E\left[x'(\bm{1}[y-d\alpha-x\beta-z\gamma<0]-\tau)\right] \label{Lagrangian} \\
&+\lambda_z \E\left[z'(\bm{1}[y-d\alpha-x\beta-z\gamma<0]-\tau)\right], \nonumber
\end{align}
where $\lambda_x$ is a $1\times k$ vector and $\lambda_z$ is a $1\times p$ vector. 
Therefore, the IVQR estimator is given by the empirical counterpart of 
\begin{equation*}
\underset{(\theta,\lambda_x,\lambda_z)}{\rm argmin}\; \mathcal{L}(\theta,\lambda_x,\lambda_z), 
\end{equation*}
where $\theta=(\alpha',\beta',\gamma')'$. 

The first derivatives of the Lagrangian in equation \eqref{Lagrangian} are
\begin{align}
\partial \mathcal{L}/\partial \alpha & =-\left\{\lambda_x
\E\left[f \cdot x'd\right] +\lambda_z \E\left[f \cdot z'd\right]\right\}' \label{s_alpha}\\
\partial \mathcal{L}/\partial \beta & = -\left\{\lambda_x \E\left[f \cdot x'x\right]
+\lambda_z \E\left[f \cdot z'x\right]\right\}' \label{s_beta}\\
\partial \mathcal{L}/\partial \gamma & = \left\{2 \gamma'A-
\lambda_x \E\left[f \cdot x'z\right] -\lambda_z \E\left[f \cdot z'z\right]\right\}' \label{s_gamma}\\
\partial \mathcal{L}/\partial \lambda_x & = \E\left[x'(\bm{1}[y-d\alpha-x\beta-z\gamma<0]-\tau)\right]' \label{s_lx}\\
\partial \mathcal{L}/\partial \lambda_z & = \E\left[z'(\bm{1}[y-d\alpha-x\beta-z\gamma<0]-\tau)\right]', \label{s_lz}
\end{align}
where $f:=f_{u_\tau}(0|d,x,z)$ denotes the density function of $u_\tau:=y-d\alpha_0(\tau)-x\beta_0(\tau)$ conditional on $s=(d,x,z)$, evaluated at the $\tau$-th conditional quantile, which is zero. Note that $f$ is specific for each quantile $\tau$. This density function plays a central role in what follows.


The solution should have all equations above equal to zero when assuming an interior solution as in Assumption \ref{A1} below. Thus, from equation \eqref{s_beta},
\begin{equation}\label{eq:aux1}
\lambda_x'=-\left(\E[f \cdot x'x]\right)^{-1}\left(\E[f \cdot x'z]\right)\lambda_z'.
\end{equation}
 Then, replacing \eqref{eq:aux1} in \eqref{s_gamma},
\begin{equation*}
\left(\E[f \cdot z'z]-\E[f \cdot z'x](\E[f \cdot x'x])^{-1}\E[f \cdot x'z]\right)\lambda_z'=2A\gamma,
\end{equation*}
such that
\begin{equation}\label{eq:aux2}
\lambda_z'=2\left(\E[f \cdot z'z]-\E[f \cdot z'x](\E[f \cdot x'x])^{-1}\E[f \cdot x'z]\right)^{-1}A\gamma.
\end{equation}

Finally, replacing \eqref{eq:aux2} in \eqref{s_alpha},
\begin{align*}
 \E\left[f \cdot d'x\right]\lambda_x'+\E\left[f \cdot d'z\right]\lambda_z' & = 
 2\left\{\E\left[f \cdot d'z\right]-\E\left[f \cdot d'x\right](\E[f \cdot x'x])^{-1}\E[f \cdot x'z]\right\}\times \\
& \left\{\E[f \cdot z'z]-\E[f \cdot z'x](\E[f \cdot x'x])^{-1}\E[f \cdot x'z]\right\}^{-1}A\gamma=\bm{0}_r,
\end{align*}
where $\bm{0}_r$ is a $r\times1$ vector of zeros.

Therefore, we can restate the IVQR problem for $(\alpha',\beta',\gamma')'$ as a system of three equations given by
\begin{align}
 & \left\{\E\left[f \cdot d'z\right]-\E\left[f \cdot d'x\right] (\E\left[f \cdot x'x\right])^{-1}\E\left[f \cdot x'z\right]\right\}\times \nonumber \\
& \left\{\E\left[f \cdot z'z\right]-\E\left[f \cdot z'x\right] (\E\left[f \cdot x'x\right])^{-1}\E\left[f \cdot x'z\right]\right\}^{-1}A\gamma=\bm{0}_r \label{g1} \\
& \E\left[x \cdot (\bm{1}[y-d\alpha-x\beta-z\gamma<0]-\tau)\right] =\bm{0}_k \label{g2} \\
& \E\left[z \cdot (\bm{1}[y-d\alpha-x\beta-z\gamma<0]-\tau)\right] =\bm{0}_p \label{g3}.
\end{align}

\subsection{First-stage IVQR parameters}

Given equations \eqref{g1}--\eqref{g3} above, we can see that \eqref{g1} provides a first-stage representation of the IVQR model. This can be written as
\begin{equation}\label{eq:g1_2}
\delta'A\gamma=\bm{0}_r,
\end{equation}
where
\begin{align}
\delta&:= \left\{\E\left[f \cdot z'z\right]-\E\left[f \cdot z'x\right] (\E\left[f \cdot x'x\right])^{-1}\E\left[f \cdot x'z\right]\right\}^{-1} \nonumber \\
&\left\{\E\left[f \cdot z'd\right]-\E\left[f \cdot z'x\right] (\E\left[f \cdot x'x\right])^{-1}\E\left[f \cdot x'd\right]\right\}.\label{eq:delta}
\end{align} 
Here $\delta$ is a $p\times r$ matrix.
Notice that equation \eqref{eq:delta} is a least-squares projection coefficient. In particular, the representation in \eqref{eq:delta} is a weighted projection, where the endogenous variable(s), $d$, is(are) regressed on the IV, $z$, and the exogenous variables, $x$. This is the analogue to the first-stage in the 2SLS case, with the difference that the QR case is a weighted regression. The weights are given by the conditional density function of the innovation term in the QR structural model, conditional on the endogeneous and exogenous covariates together with the instruments.

Hence, for each endogeneous variable, say $d_j$ for $j=1,2,...,r$, $\delta_{j}$ in equation \eqref{eq:delta} can be recovered as the solution to the following optimization problem
\begin{equation}\label{eq:first stage}
\mu_j:=(\psi_j,\delta_j)=\underset{\psi,\delta}{\rm argmin} \;\E \left[f \cdot (d_j-x\psi- z\delta)^2\right].
\end{equation}

Note that the parameter $\delta$ also depends on $\theta=(\alpha',\beta',\gamma')'$, through the conditional density function $f$ at quantile $\tau$. Thus, this first-stage representation depends on the structural (second-stage) parameters, and as such, it is different from the 2SLS case in mean regression models.

We notice that the first-stage in equation \eqref{eq:first stage} is different from those in the existing literature using two-stage regressions for conditional quantile models. \citet{Amemiya82}, \citet{Powell83}, \citet{ChenPortnoy96}, and \citet{KimMuller04} propose different two step procedures in which the first step fits the endogenous variable(s) as a function of exogenous covariates and IV, and this is then plugged in a second-stage. Nevertheless, these papers use least squares without weighting or standard quantile regression in the first-stage. Our procedure derives the first-stage from the IVQR set-up, thus confirming that  a first-stage (albeit different) is part of the model.

\subsection{Relation to the Jacobian condition}

Now we consider the relationship between the first-stage derived in the previous section, in particular equation \eqref{eq:delta}, and the rank identification conditions for the IVQR estimator of CH. For simplification, we consider a model without additional exogenous covariates $x$. 

As discussed in CH, the IVQR optimization problem  is asymptotically equivalent to solving the following moment condition  
\begin{equation*}	
\Pi((\alpha,\gamma),\tau)=\E\left[z'(\bm{1}[y-d\alpha-z\gamma<0]-\tau)\right]=\bm 0_p.
\end{equation*}
To establish the asymptotic properties of the IVQR estimator, it is required that the Jacobian matrices, $\frac{\partial}{\partial\alpha}\Pi((\alpha,\gamma),\tau)$ and $\frac{\partial}{\partial\gamma}\Pi((\alpha,\gamma),\tau)$, are continuous and full column rank (see below the conditions for the derivation of the asymptotic properties of the estimator, in particular, Assumption 1, item R3). We show here that these conditions are embedded in the FS-IVQR representation.
	
The rank Jacobian conditions are
\begin{align*}
\textnormal{rank}\left(\frac{\partial\Pi((\alpha,\gamma),\tau)}{\partial\alpha}\right) & = \textnormal{rank}\left(\E[f\cdot  z'd]\right)\geq r,\\
\textnormal{rank}\left(\frac{\partial\Pi((\alpha,\gamma),\tau)}{\partial\gamma}\right) & =\textnormal{rank}\left(\E[f\cdot  z'z]\right)=p.
\end{align*}
The first equation implies that for the case of one endogeneous variable, $r=1$, $\E[f\cdot  z'd]$ has at least one non-zero column, and the second equation requires $p$ noncollinear valid instruments. Now notice that from the FS-IVQR representation given by equation \eqref{eq:delta}, in the  case without exogenous regressors, simplifies to: 
\begin{equation*}
\E[f\cdot z'd]=\E[f\cdot z'z]\delta. 
\end{equation*}
Therefore, the matrices involved in the rank conditions directly appear in representation \eqref{eq:delta}. Also, if the FS-IVQR parameter $\delta=0$, then the rank conditions cannot be satisfied. Note that this is a necessary condition, but not a sufficient one. Furthermore, by checking how close $\delta$ is to zero one is in fact evaluating the strength of the identification condition.

\subsection{Further intuition on the FS-IVQR}

The restriction in equation \eqref{eq:g1_2} provides a natural framework to evaluate the relevance of the instruments in IVQR models. 

First, the first-stage regression representation in \eqref{eq:first stage} is a weighted linear projection, where the weights are the conditional density function of the innovation term in the QR structural model, conditional on the endogeneous and exogenous covariates together with the instruments. This exposes a caveat of the QR IV model. In order to estimate the parameters in  \eqref{eq:first stage} consistently, one needs a consistent estimate of the density $f$, and hence at least one valid instrument must be available to the researcher. This is in contrast with the standard conditional average models where the first-stage is a simple OLS regression without weights. The required weights in the QR case will be further discussed below when we suggest a test for the validity of the IV.

Second, notice that the parameter $\delta$ captures the strength of the instrument in the sense it measures the correlation between the instrument $z$ and the endogenous variable $d$ weighted by the density function $f$. This is the QR counterpart of the first-stage partial correlation of $z$ on the endogenous variables $d$ for the 2SLS. As noted by \citet{GalvaoMontes15} the CH set-up is equivalent to the 2SLS in least-squares models. In fact the CH estimator is the QR counterpart of a 2SLS estimator. The expression above also shows that there is an implicit first-stage, similar to that in 2SLS problems. As such, this provides an analytical expression to evaluate the relevance of the IV. When the instrument is valid, $\delta\neq\bm{0}_{p\times r}$. 

Third, note that the instrument $z$ does not belong in the structural quantile model \eqref{eq:structural}, hence  $\gamma=\bm{0}_{p\times r}$ can be used for identification, a key feature of the CH IVQR estimator.  Equation \eqref{eq:g1_2} also shows that when $\delta=\bm{0}_{p\times r}$, the value of $\gamma$ is irrelevant, and therefore it cannot be used in the IVQR procedure to solve endogeneity. As such, $\delta\neq\bm{0}_{p\times r}$ is a necessary condition for the IV to have a purpose in the CH set-up. Therefore, a test for the validity of the instruments can be based on a test for statistical significance of $\delta$.

Finally, another way of gaining intuition on the test is the following. Assume that $r=1$ (i.e. only one endogenous variable), then \eqref{eq:g1_2} is in fact equal to 0, a scalar. If we further assume that $A=I_p$, then
\begin{equation}
\sum_{q=1}^{p}\delta_q\gamma_q=0,
\end{equation}
where $\delta=[\delta_1,\ldots,\delta_p]'$ is the column vector that has the first-stage effect of all IV on $d$. Note again that if $\delta=\bm{0}_{p\times 1}$, then the vector $\gamma$ could have any value and its implied restrictions would be irrelevant.

\section{Formulation of the test for validity of the IV}\label{test}

In this section we suggest tests for the validity of the IV using the first-stage representation. The formulation of the test proposed in this paper is based on the condition given in equation \eqref{g1} together with the first-stage IVQR representation in equation \eqref{eq:delta}. A test for validity of the instruments for $p$ instruments can be based on the null hypothesis
\begin{equation}\label{eq:null}
H_0:\delta_0=\bm{0}_{p\times r},
\end{equation}
against the alternative
\begin{equation}\label{eq:alternative}
H_A:\delta_0\neq \bm{0}_{p\times r}.
\end{equation}

We highlight that, differently from the 2SLS, the first-stage IVQR in \eqref{eq:first stage} is for a given quantile $\tau$. Thus, for the same variables $d$ and instruments $z$, the strength of the instruments may vary across different quantiles. This variation is captured by the weights $f$.

Note that the procedure works for $r\geq 1$, that is for one or more than one endogenous variables. In the $r>1$ case, separate tests could be applied as in 2SLS analysis where there may be a different first-stage for each endogeneous variable. To simplify the procedures below we assume that $r=1$, that is, there is only one endogenous variable.

The expressions of the null and the alternative hypotheses in \eqref{eq:null} and \eqref{eq:alternative}, respectively, lead to the following testing procedure.

When $H_0$ is true, under suitable regularity conditions, $\hat{\delta}$ converges in probability to $\bm{0}_{p\times r}$ for a given $\tau$. On the other hand, when $H_A$ is true, $\hat{\delta} $ 
converges in probability to $\delta_0\neq \bm{0}_{p\times r}$. Therefore, it is reasonable
to reject $H_0$ if the magnitude of $\hat{\delta} $ is suitably large. 

A natural choice to test $H_0$ against $H_1$ for the case of $r=1$ is the Wald statistic as
\begin{equation}\label{eq:test}
T_n=n \hat\delta ' \{ V_{\delta} \}^{-1} \hat\delta , 
\end{equation}
where $V_{\delta}$ is the asymptotic covariance matrix of $\sqrt{n} \hat{\delta}$ under $H_0$. In practice, $V_{\delta}$  is replaced by a suitable consistent estimate. We will discuss the practical implementation as well the limiting distribution in the next section.

\section{Empirical implementation and asymptotic distribution}\label{estimator.and.asymptotics}

In this section we propose a two step estimator for the first-stage instrumental variables quantile regression (FS-IVQR), consider its empirical implementation, and derive the estimators' asymptotic distribution. The two steps estimation procedure consists of estimating the conditional density using the IVQR model in the first step, and in the second step employing a weighted least squares (WLS) regression. For simplicity of exposition, we present the case of $r=1$, i.e. one endogenous variable, but as discussed above the case of $r>1$ can be implemented using separate regressions.

\subsection{FS-IVQR Estimator}\label{estimator}

The FS-IVQR estimator requires a consistent estimator of $\mu$ in \eqref{eq:first stage}, which will be based on WLS based on the estimator of $f$, at a given quantile of interest $\tau$. The estimator has two steps as following:

\vspace{0.25cm}

\noindent \textbf{1)} In the first step we obtain $\hat{\theta}=(\hat{\alpha}, \hat{\beta}', \hat{\gamma}')'$ from the CH estimator, 
\begin{equation*}
\hat{\alpha}=\underset{\alpha}{{\rm argmin}}{\|\hat\gamma(\alpha)\|_{A}},
\end{equation*}
where
\begin{equation*}
(\hat{\beta}(\alpha),\hat{\gamma}(\alpha))=\underset{\beta,\gamma}{\rm argmin}\; \frac{1}{n} \sum_{i=1}^{n} \left[\rho_\tau(y_{i}-d_{i}\alpha-x_{i}\beta-z_{i}\gamma) \right].
\end{equation*}

Provided that the $\tau$th conditional quantile function of $y|s$ is linear, as in \eqref{eq:structural}, then for $h_{n}\to 0$ we can consistently estimate the parameters of the $\tau \pm h_{n}$ conditional quantile functions by $\hat{\theta}(\tau \pm h_{n})$. And the density $f_{i}:=f_{u_\tau}(0|d=d_{i},x=x_{i},z=z_{i})$ can thus be estimated by the difference quotient
\begin{equation}\label{eq:sparsity}
\hat{f}_{i}=\frac{2h_{n}}{s_{i}\left( \hat{\theta}(\tau+h_{n})-\hat{\theta}(\tau-h_{n}) \right)}.
\end{equation}
 The estimation in \eqref{eq:sparsity} is a natural extension of sparsity estimation methods, suggested by \citet{HendricksKoenker92}.\footnote{We note that a kernel estimator for the conditional density, as in \citet{Powell91}, can be used. The procedure would use the error term $\hat{u}_\tau:=y-d\hat{\alpha}(\tau)-x\hat{\beta}(\tau)$ from the first step CH IVQR estimator. We describe the procedure using the sparsity estimation for simplicity.} The estimator is discussed in further details in \citet{ZhouPortnoy96} and \citet{Koenker05}.
We introduce the simplifying notation $\hat{f}_{i}:=\hat{f}_{u_\tau}(0|s=s_i)$.\footnote{We are assuming that there is only one endogenous variable, $r=1$. Otherwise the analysis below should be repeated separately for each endogenous variable as there will be a different first-stage for each one.} The bandwidth for the density estimation can be chosen heuristically as a scaled version of \citet{HallSheather88}:
\begin{equation*}
h_n=2n^{-1 / 3} \Phi^{-1}\left(0.975\right)^{2 / 3}\left[\frac{3}{2} \cdot \frac{\phi\left\{\Phi^{-1}(\tau)\right\}^{4}}{2 \Phi^{-1}(\tau)^{2}+1}\right]^{1 / 3}.
\end{equation*}

\vspace{0.25cm}

\noindent \textbf{2)} In the second step the parameters of interest $\delta$ can be obtained from a feasible WLS as
\begin{equation}\label{eq:wls}
\hat{\mu}:=(\hat{\psi},\hat{\delta})=\underset{\psi,\delta}{\rm argmin} \; \frac{1}{n} \sum_{i=1}^{n} \left[\hat{f}_{i}\cdot (d_{i}-x_{i}\psi- z_{i}\delta)^2 \right].
\end{equation}
Equation \eqref{eq:wls} produces $\hat{\delta}$ which is the main object of interest. 

Define $Y$, $X$, $D$  and $Z$ as the matrices formed from a random sample of $\{y_i,d_i,x_i,z_i\}_{i=1}^n$. Similarly define $W=[X,Z]$. Define the weighting diagonal matrix 

\begin{equation*}
\hat{V} = 
\begin{bmatrix}
\hat{f}_{1} &  &  \\
& \ddots &  \\
&  & \hat{f}_{n}
\end{bmatrix}.
\end{equation*}

Then, the estimator in \eqref{eq:wls} above can be written in a simple matrix notation as
\begin{equation}\label{eq:wls_mu}
\hat{\mu}=(W' \hat{V}  W)^{-1}W' \hat{V} D.
\end{equation}

Notice that if $f_i$ is a constant for all $i$, then the proposed FS-IVQR method should deliver same estimates as FS-2SLS for the mean. This would happen, for example, in the case of $i.i.d.$ innovations in the second-stage structural model. Thus, there will be differences between the two estimators only when $f_i$ varies across $i$, that is, when the weighting factor is not a constant. Example 1 (location model)  Appendix B shows a case where the density function is a constant. A typical example where the weights are not constant across individuals is the  location-scale model, see Examples 2 and 3 in Appendix B.

\subsection{Asymptotic distribution}\label{asymptotics}

In this subsection, we derive the asymptotic distribution of the proposed estimator. The asymptotic properties of the IVQR estimator can be found in \citet{CH06} and the assumptions therein are those required for inference. We consider Assumption 2 in \citet[][pp.501--502]{CH06}, that we reproduce here for convenience. It imposes conditions for $\theta_0$ to be identified and estimated.
\begin{assumption}\label{A1}
R1. Sampling. $\{y_i,x_i,d_i,z_i\}$ are $iid$ defined on a probability space and take values in a compact set.\\
R2. Compactness and convexity. For all $\tau\in(0,1)$, $(\alpha,\beta,\gamma\in {\rm int} (\mathcal{A}\times\mathcal{B}\times\mathcal{G})$ is compact and convex.\\
R3. Full rank and continuity. $y$ has bounded conditional density (conditional on $w$), and for $\theta=(\alpha,\beta,\gamma)$,
\[\Pi(\theta,\tau):=\E\left[(\tau-\bm{1}(y<d\alpha+x\beta+z\gamma)\cdot[x,z]\right],\]
Jacobian matrices $\frac{\partial}{\partial(\alpha',\beta')}\Pi(\theta,\tau)$ and $\frac{\partial}{\partial(\beta',\gamma')}\Pi(\theta,\tau)$  are continuous and have full rank, uniformly over $\mathcal{A}\times\mathcal{B}\times\mathcal{G}$ and the image of $\mathcal{A}\times\mathcal{B}\times\mathcal{G}$ under the mapping $(\alpha,\beta)\mapsto\Pi(\theta,\tau)$ is simply connected. Assume that $\theta_0=(\alpha_0,\beta_0',\gamma_0')'$ is the unique solution to the CH problem.
\end{assumption}


We impose additional conditions for deriving the limiting properties of the feasible first-stage estimator in \eqref{eq:wls} using the sparsity estimation in \eqref{eq:sparsity}. 

\begin{assumption}\label{A2}
Let $\varepsilon_{i}:=d_{i}-x_{i}\psi_0- z_{i}\delta_0$, with $\E[\varepsilon_{i} | w_{i} ] = 0$, and $\E[\varepsilon_{i}^{2}|w_{i}] = \sigma_{i}^2$.  Also, let $f_{i}:=f_{\theta_0}(y-s\theta_0|s=s_i)$ and assume that $\E[|f_{i}^{-2}w_{i} \varepsilon_{i}|]<\infty$. Let $\Omega_{f \sigma}:=\E[f_{i}^2\sigma_{i}^{2} w_{i} w_{i}']$ and $\Omega_{f }:=\E[f_{i} w_{i} w_{i}']$. The limits $\lim_{n\to\infty}\frac{1}{n}\sum_{i}^{n} f_{i}^2\sigma_{i}^{2} w_{i} w_{i}' = \Omega_{f \sigma}$ and $\lim_{n\to\infty}\frac{1}{n}\sum_{i}^{n} f_{i} w_{i} w_{i}' = \Omega_{f }$ exist and are nonsingular (and hence finite).
\end{assumption}

Assumption \ref{A2} contains conditions for establishing consistency and asymptotic normality of the proposed estimator. The next result presents an intermediate result.

\begin{lemma}\label{lemmaJ}
Under Assumptions \ref{A1}--\ref{A2}, as $n\rightarrow\infty$, $h_n\rightarrow0$ and $nh_n^2\rightarrow\infty$, 
\begin{equation}
\sqrt{n}\left( \hat{\mu}- \mu_0 \right) \stackrel{d}{\rightarrow} \mathcal{N}\left(\bm{0}_{k+p},  V(\mu_0) \right),
\end{equation}
where $\mu_0:=(\psi_0,\delta_0)=\underset{\psi,\delta}{\rm argmin} \;\E \left[f \cdot (d-x\psi- z\delta)^2\right]$ and
$V(\mu_0) = \Omega_{f}^{-1} \Omega_{f \sigma} \Omega_{f}^{-1}$ is the asymptotic covariance matrix. 
\end{lemma}
\begin{proof}
	In Appendix A.
\end{proof}


\subsection{Asymptotic distribution of the test statistic}\label{sec: asymp test}

Consider a subset of the instruments, $p_1<p$, and consider a partition of $\delta=[\delta_1',\delta_2']'$ of the corresponding first-stage parameters of interest, with dimensions $p_1$ and $p_2$ (with $p=p_1+p_2$), respectively. Consider a $p_1\times (k+p)$ matrix $R=[\bm{0}_{p_1\times k}, \bm{I}_{p_1},\bm{0}_{p_1\times p_2}]$ where $\bm{I}_{p_1}$ is an identity matrix of dimension $ p_1\times p_1$. Thus, $R\mu=\delta_1$ is the subvector of interest. Let $\hat{V}(\hat{\mu})$ be a consistent estimator of $V(\mu_0)$, which can be obtained from the WLS procedure. The next result derives the limiting distribution of the test statistic in equation \eqref{eq:test}.

\begin{proposition}\label{prop:delta}
	Consider Assumptions \ref{A1}--\ref{A2}, $n\rightarrow\infty$, $h_n\rightarrow0$ and $nh_n^2\rightarrow\infty$. Furthermore, assume that $dim(z)=p>p_1\geq 1$. Then, under $\delta_{2}\neq\bm{0}_{p_2}$ and $H_0:\delta_{1}=\bm{0}_{p_1}$ and local alternatives $H_A:\delta_{1}=\bm{a}_{p_1}/\sqrt{n}$, 
\begin{equation}\label{T_n}
	T_n=n \left(R \hat{\mu} \right)' \{R\hat{V}(\hat{\mu})R'\}^{-1} \left(R \hat{\mu} \right)
	\stackrel{d}{\rightarrow} \chi^2_{p_1}(\bm{a}_{p_1}).
\end{equation}
\end{proposition}
\begin{proof}
	In Appendix A.
\end{proof}

Computation of the test statistic \eqref{eq:test} requires a non-parametric estimator of $f$, the conditional density of $u_\tau|d,x,z$ evaluated at the specific quantile of interest $\tau$.
Given that the weights need to be estimated, the proposed FS-IVQR has specific properties when testing under the null hypothesis of an invalid instrument. The condition on the number of IV being larger than the number of parameters tested in the null hypothesis is required for consistent estimation of $\theta$ under the null, which in turn, is used for the consistent estimation of $f$. 

\begin{remark}
As noted above, testing for the null \eqref{eq:null} that the instruments are \textit{not valid} requires a consistent estimate of the density $f$, which in turn requires at least one valid instrument, and hence the restrictions on the availability of at least one valid instrument as well as on the number of IV being larger than the number of parameters tested in the null hypothesis. Hence, due to the required estimation of weights in the first stage, it is difficult to establish a complete analogous F-statistic type rule-of-thumb for categorizing weak instruments as in \citet{Staiger97}, and subsequent variants as \citet{Sanderson16}, \citet{Lee20} and others for OLS models (see \citet{StockYogo05} for an extensive discussion). Nevertheless, dropping the multiple instruments requirement is a very interesting line of future research for investigating the weak instruments problem. Therefore, we suggest to practitioners to use the methods for the QR first-stage together with the robust inference procedures for weak identification in QR models proposed in \citet{CH08}, \citet{Jun08}, \citet{CHJ09} and \cite{AndrewsMikusheva16}.
\end{remark}

\section{Monte Carlo experiments}\label{MonteCarlo}

We analyze in this section the performance of the proposed test with finite samples through a series of Monte Carlo simulation exercises. 
The data generating process (DGP) has the following location-scale model:
\begin{align}
y_i &= d_i + x_i + (1+d_i)u_i, \label{eq:MCstructural}\\
d_i &= c_1 + a z_{1i} + \phi z_{2i} + (1 + b z_{1i}) v_i, \label{eq:MCfirststage}
\end{align}
where $x_i$, $z_{1i}$ and $z_{2i}$ are three independent variables with distribution $U(0,1)$; $u_i$ and $v_i$ have standard bivariate normal distribution with correlation $0.50$. The constant parameter $c_1=10$ is set to a large value to satisfy the monotonicity assumption. Equations \eqref{eq:MCstructural}--\eqref{eq:MCfirststage} specify a model where there could be pure location or location-scale specifications in the first stage, thus allowing the instruments to have different effects on the endogenous variable.
Note that the parameters $a$ and $b$ determine the type of effect that the instrument $z_{1}$ has on the endogenous covariate $d$. For example, if $a\neq0$ and $b = 0$ the instrument $z_{1}$ has a pure location effect on $d$ (pure location shift model), while if $a = 0$ and $b\neq0$ the effect is only on the variance of the endogenous covariate (pure scale shift model). 

In all cases we consider tests for $H_0:\delta_1=0$ where this is the first-stage parameter associated with the $z_1$ instrument defined in the previous sections. 
We consider two different cases to investigate the numerical properties of the tests. In the first case, $\phi=1$, there is a second instrument, $z_{2}$, such that the model correctly identifies the parameters in the structural equation \eqref{eq:MCstructural} for all possible values of $a$ and $b$, even under the case that $a=b=0$. In the second case, we set $\phi=0$, and therefore, under the null hypothesis the consistent estimation of the weights $f$ is problematic. Also, in this case, when $a=b=0$, there is no valid available instrument.

We will consider three different test statistics from different estimators. First, for comparison purposes, we present a Wald test for the coefficient in $z_1$ using a simple regression model of $d$ on $(x,z_1,z_2)$ in a standard 2SLS framework, denoted FS-2SLS. Second, we test for  $H_0:\delta_1=0$ using the true density function, $f$, as weights, that is, using the true $\theta_0$, denoted FS-IVQR (true density). We note that this is not observed in practice, and we include these results for comparison purposes. Our proposed test studied in the previous section is the third one, denoted FS-IVQR (sparsity), 
where we use the sparsity function estimation described above. Note that the three tests differ only in the weighting procedure used in the regression of $d$ on $(x,z_1,z_2)$ or $(x,z_1)$.

Table \ref{table:sizec1psi1}--\ref{table:sizec1psi0} show the empirical size (i.e. $a=b=0$) of the computed test with 2000 simulations for $ n = \{500, 1000 \}$ and for the quantiles $\tau = \{0.25, 0.50, 0.75\}$. 

Consider first the case where there is a second instrument, $\phi=1$ in Table \ref{table:sizec1psi1}. The tests have approximately correct empirical size in all cases. As such they clearly evaluate if the instrument $z_1$ exerts an effect on the endogenous variable $d$. In all cases they have a similar performance to the FS-2SLS case. 

Now consider the case where there is no available second instrument, $\phi=0$ in Table \ref{table:sizec1psi0}. The idea of this experiment is to evaluate the test performance when there is lack of identification under the null. In this case, the weights in the structural model cannot be estimated consistently under the null. Since the proposed test evaluates the relationship between $z_1$ and $d$, the main issue is whether this relationship can be evaluated in other than the OLS model. The simulations show that the size is correct for the sparsity estimator. This result suggests that the test can be used even when the structural parameters cannot be estimated under the null (because $z_1$ does not solve the endogeneity problem). 

This is not a general result, however, but it illustrates the role of the density function as a weighting factor.
In order to explore this, we consider three examples in Appendix B closely related to the DGP used in the Monte Carlo experiments. 
When the instrument $z$ is available we should be estimating the correct structural parameters and $f_{u_\tau}(0|d,x,z)$ where $u_\tau=y-Q_\tau(y|d,x,z)$. However, the case where, under the null, $z$ is invalid would be equivalent to the case where there are no instruments available. That is, we would not be able to solve the endogeneity in the second-stage. Note that, for this case, the density function that will be implicitly used is that of $u^*_\tau=y-Q_\tau(y|d,x)$. The examples in Appendix B compare  $f_{u^*_\tau}(0|d,x)$ with  $f_{u_\tau}(0|d,x,z)$.\footnote{Let  $\tilde{\alpha}$ and $\tilde{\beta}$ be the parameters that result from the estimation of the biased structural model without instruments, $Q_\tau(y|d,x)=d\tilde{\alpha}+x\tilde{\beta}$.  Note that $u_\tau=y-Q_\tau(y|d,x,z)=y-d\alpha_0-x\beta_0$ can be written as $y-d\tilde\alpha-x\tilde\beta-bias(d,x)$, where $bias(d,x)=d(\alpha_0-\tilde{\alpha})+x(\beta_0-\tilde{\beta})$ such that
	$u_\tau=u^*_\tau-bias(d,x)$.} The partial results suggest that if $f_{u_\tau}(0|d,x,z)$ and $f_{u^*_\tau}(0|d,x)$ are proportional to each other when they vary with $d$, we could implement the first-stage test under the null of all IV being invalid.

\begin{table}[htbp]
  \centering
  \caption{Rejection rate of the null hypothesis using $a=b=0$ and $\phi=1$}
  \label{table:sizec1psi1}
\begin{tabular}{ccccccccc}
\hline
\multicolumn{ 1}{c}{{\it {\boldmath$\tau$}}} & \multicolumn{ 1}{c}{{\it {Size}}} &   \multicolumn{ 3}{c}{{\boldmath $n = 500$}} &     {\bf } &  \multicolumn{ 3}{c}{{\boldmath $n = 1000$}} \\

\multicolumn{ 1}{c}{{\it {\bf }}} & \multicolumn{ 1}{c}{{\it {\bf }}} &  {\bf FS-2SLS} & {\bf True \boldmath $f$} & {\bf Sparsity \boldmath $f$} &     {\bf } &  {\bf FS-2SLS} & {\bf True \boldmath $f$} & {\bf Sparsity \boldmath $f$} \\

\hline
\multicolumn{ 1}{c}{{\bf 0.25}} & {\bf 0.10} &      0.100 &      0.104 &      0.101 &            &      0.104 &      0.105 &      0.108 \\

\multicolumn{ 1}{c}{{\bf }} & {\bf 0.05} &      0.052 &      0.051 &      0.053 &            &      0.054 &      0.056 &      0.054 \\

\multicolumn{ 1}{c}{{\bf }} & {\bf 0.01} &      0.014 &      0.015 &      0.016 &            &      0.008 &      0.009 &      0.010 \\
\hline
\multicolumn{ 1}{c}{{\bf 0.50}} & {\bf 0.10} &      0.100 &      0.104 &      0.109 &            &      0.096 &      0.095 &      0.095 \\

\multicolumn{ 1}{c}{{\bf }} & {\bf 0.05} &      0.056 &      0.055 &      0.059 &            &      0.056 &      0.056 &      0.053 \\

\multicolumn{ 1}{c}{{\bf }} & {\bf 0.01} &      0.011 &      0.011 &      0.010 &            &      0.011 &      0.013 &      0.012 \\
\hline
\multicolumn{ 1}{c}{{\bf 0.75}} & {\bf 0.10} &      0.097 &      0.103 &      0.101 &            &      0.102 &      0.104 &      0.100 \\

\multicolumn{ 1}{c}{{\bf }} & {\bf 0.05} &      0.047 &      0.051 &      0.049 &            &      0.054 &      0.054 &      0.058 \\

\multicolumn{ 1}{c}{{\bf }} & {\bf 0.01} &      0.011 &      0.009 &      0.013 &            &      0.009 &      0.010 &      0.012 \\
\hline
\end{tabular}  
  \label{tab:addlabel}
  
  Note: Rejection rates of 2000 Monte Carlo experiments.
\end{table}

\begin{table}[htbp]
	\centering
	\caption{Rejection rate of the null hypothesis using $a=b=0$ and $\phi=0$}
	\label{table:sizec1psi0}
\begin{tabular}{ccccccccc}
\hline
\multicolumn{ 1}{c}{{\it {\boldmath$\tau$}}} & \multicolumn{ 1}{c}{{\it {Size}}} &   \multicolumn{ 3}{c}{{\boldmath $n = 500$}} &     {\bf } &  \multicolumn{ 3}{c}{{\boldmath $n = 1000$}} \\

\multicolumn{ 1}{c}{{\it {\bf }}} & \multicolumn{ 1}{c}{{\it {\bf }}} &  {\bf FS-2SLS} & {\bf True \boldmath $f$} & {\bf Sparsity \boldmath $f$} &     {\bf } &  {\bf FS-2SLS} & {\bf True \boldmath $f$} & {\bf Sparsity \boldmath $f$} \\

\hline
\multicolumn{ 1}{c}{{\bf 0.25}} & {\bf 0.10} &      0.110 &      0.114 &      0.107 &            &      0.093 &      0.097 &      0.101 \\

\multicolumn{ 1}{c}{{\bf }} & {\bf 0.05} &      0.057 &      0.060 &      0.062 &            &      0.054 &      0.057 &      0.054 \\

\multicolumn{ 1}{c}{{\bf }} & {\bf 0.01} &      0.013 &      0.012 &      0.016 &            &      0.013 &      0.013 &      0.010 \\
\hline
\multicolumn{ 1}{c}{{\bf 0.50}} & {\bf 0.10} &      0.107 &      0.107 &      0.108 &            &      0.097 &      0.097 &      0.094 \\

\multicolumn{ 1}{c}{{\bf }} & {\bf 0.05} &      0.059 &      0.059 &      0.059 &            &      0.049 &      0.050 &      0.051 \\

\multicolumn{ 1}{c}{{\bf }} & {\bf 0.01} &      0.017 &      0.018 &      0.020 &            &      0.010 &      0.009 &      0.009 \\
\hline
\multicolumn{ 1}{c}{{\bf 0.75}} & {\bf 0.10} &      0.101 &      0.100 &      0.108 &            &      0.102 &      0.109 &      0.105 \\

\multicolumn{ 1}{c}{{\bf }} & {\bf 0.05} &      0.049 &      0.050 &      0.053 &            &      0.054 &      0.051 &      0.052 \\

\multicolumn{ 1}{c}{{\bf }} & {\bf 0.01} &      0.010 &      0.010 &      0.009 &            &      0.008 &      0.009 &      0.011 \\
\hline
\end{tabular}  
	\label{tab:addlabel}
	
	Note: Rejection rates of 2000 Monte Carlo experiments.
\end{table}

\begin{figure}
	\centering
	\caption{Power for $H_0:\delta_1=0$ (model with $\phi=1$)}
	\includegraphics[width=12cm]{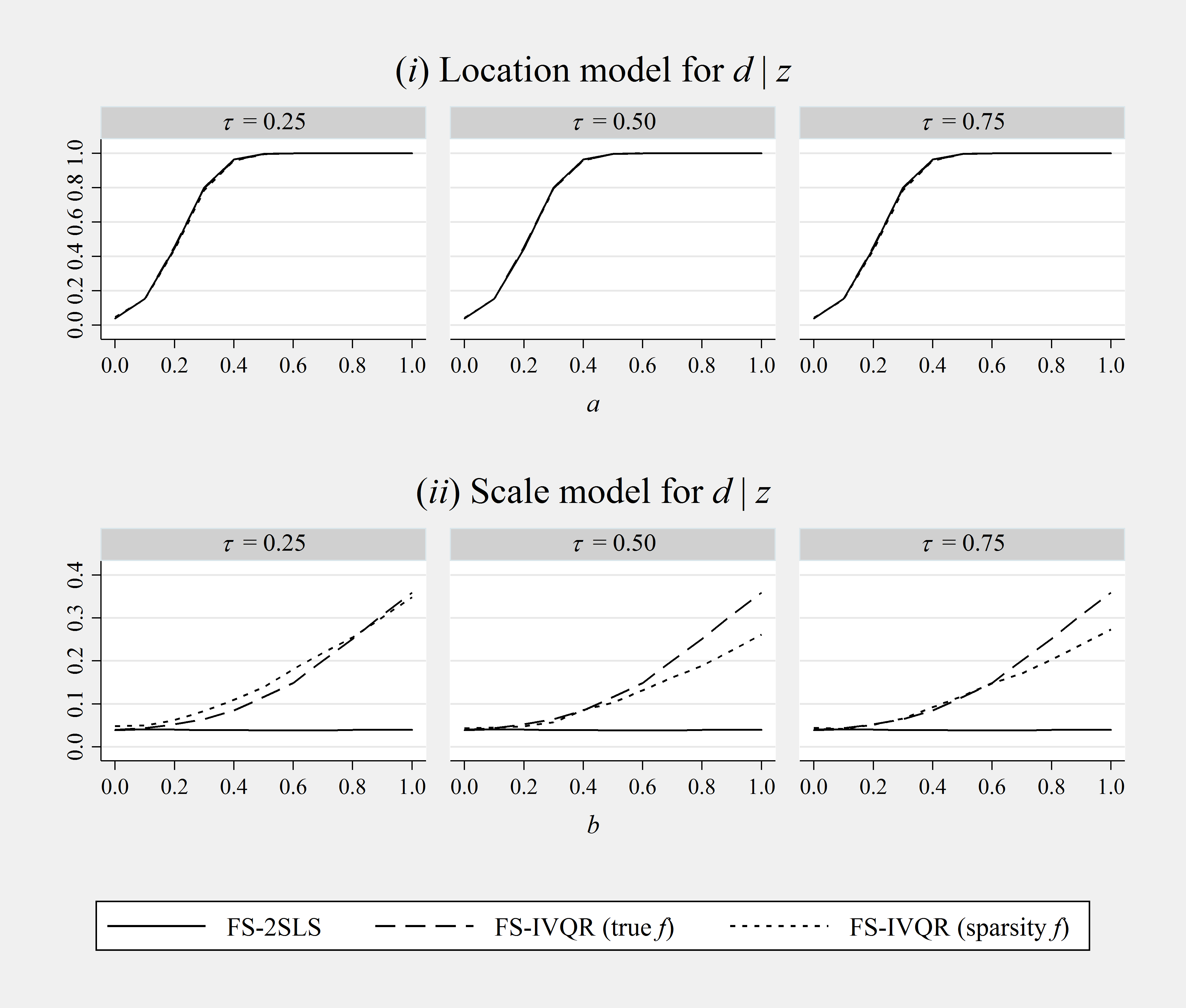} 
	\label{fig:Fig_c1_psi1}
\end{figure}

\begin{figure}
	\centering
	\caption{Power for  for $H_0:\delta_1=0$ (model with $\phi=0$)}
	\includegraphics[width=12cm]{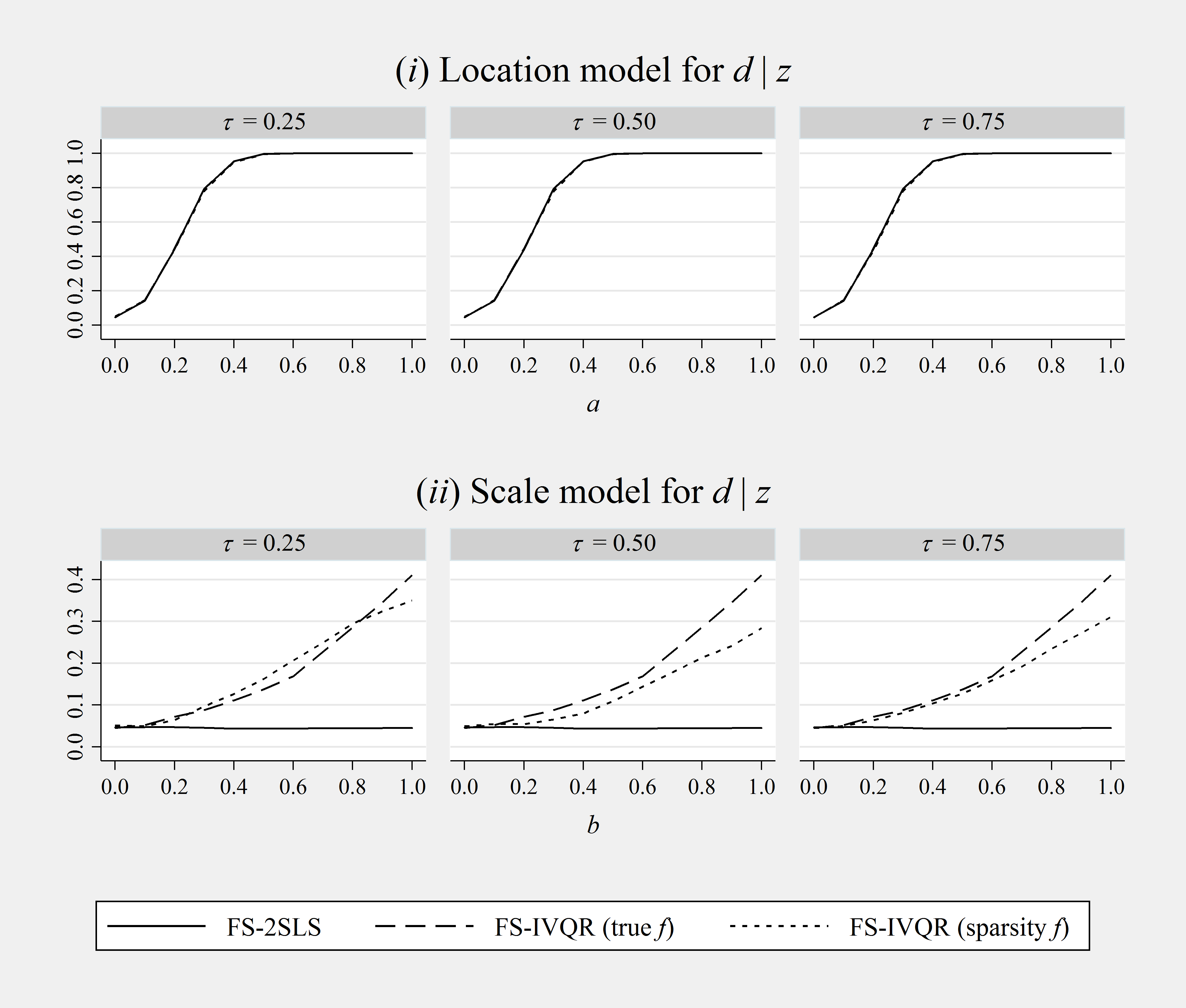} 
	\label{fig:Fig_c1_psi0}
\end{figure}

To analyze the empirical power of the tests, we performed 2000 simulations only for the case with $n = 1000$ and we calculated the rejection rates of the proposed procedure for the quantiles $\tau = \{0.25, 0.50, 0.75\}$. As benchmark we also use the test rejection rates obtained in the FS-2SLS method, i.e., the Wald test of an OLS regression of $d$ on $z_1$. The results appear in Figures \ref{fig:Fig_c1_psi1} and \ref{fig:Fig_c1_psi0}. For each figure we have two blocks, (i) and (ii), where in (i) we evaluate a pure location first-stage model of $z_1$ on $d$ using $a = \{0, 0.10, ..., 0.90,1\}$ and $b=0$, and in (ii)  we set $a=0$ and we vary $b=\{0, 0.10, ..., 0.90,1\}$ such that $z_1$ has only a scale effect on $d$.

We first consider the case where there is a second valid instrument $\phi=1$. 
Figure \ref{fig:Fig_c1_psi1}, block (i) pure location first-stage, shows that the FS-IVQR power computed with true and estimated densities behaves similarly to FS-2SLS. That is, they correctly reject as $a$ increases. The estimated density model has slightly less power than the one with the true density.  For block (ii),  the results of the FS-IVQR differ when we are in the presence of a pure-scale model for $d|z_1$. Note that in this case there is no relationship between $d$ and $z_1$ at the mean (FS-2SLS), but it does affect the other points of the conditional distribution. Therefore, the first-stage of 2SLS does not find any relationship between the endogenous variable and the instrument while the FS-IVQR estimators (both true and estimated weights) are able to correctly detect it. 

Finally, consider the last case when $\phi=0$ in Figure \ref{fig:Fig_c1_psi0}. The FS-IVQR tests also work in this case. In both (i) and (ii) cases, the tests detect an association between the instrument and the endogenous variable. In case (ii) the FS-IVQR rejects as $b$ increases while FS-2SLS does not. As noted in Table \ref{table:sizec1psi0} the test works even for the case where $a=b=0$ and the endogeneity problem in the structural estimators cannot be solved.

\section{Empirical application: Card (1995) college proximity as an instrument for education}\label{empirical}

In this section we show an application of the proposed test to a Mincer equation to estimate returns to schooling. The data used is taken from \citet{Card95} and correspond to 3010 individuals of the US National Longitudinal Survey of Young Men.\footnote{Downloaded from \url{http://davidcard.berkeley.edu/data_sets/proximity.zip}} Following the same specification of that paper, the model describes wages as a function of the years of education and other exogenous controls such as work experience, race and a set of geographic and regional variables. A classic problem with this model is that ability is unobservable and therefore its omission induces a potential bias due to endogeneity of the OLS estimator. Specification errors have analogous consequences on QR estimators, as analyzed by \citet{Angrist06}. \citet{Card95} proposes to implement an IV strategy using two measures of proximity to the university as external variables to the wage equation. For this application we set $A$, the weighting matrix in the CH-IVQR estimator, equal to the inverse of the asymptotic covariance matrix of $\hat{\gamma}$, as suggested by CH. The grid used in the minimization problem is $\alpha(\tau) = \{0,0.004,0.008,...,0.992,0.996,1\}$.

\begin{table}[htbp]
	\centering
	\caption{Returns to schooling (Card, 1995)}\label{table:card}
	\scriptsize
	\begin{tabular}{rcccc}
		\hline
		{\bf } & \multicolumn{ 1}{c}{{\bf 2SLS}} & \multicolumn{ 3}{c}{{\bf IV Quantile Regression}} \\
		
		{\bf } & \multicolumn{ 1}{c}{{\bf }} &   {\bf $\tau=0.25$} &   {\bf $\tau=0.50$} &   {\bf $\tau=0.75$} \\
		\hline
		{\bf } &  \multicolumn{ 4}{c}{{\bf First-stage estimates}} \\
		\hline
		Lived Near 2-year College in 1966 &      0.123 &     0.0644 &   0.471*** &    0.154** \\
		
		&   (0.0774) &    (0.129) &   (0.0704) &   (0.0709) \\
		
		Lived Near 4-year College in 1966 &   0.321*** &   0.380*** &   0.298*** &     0.140* \\
		
		&   (0.0878) &    (0.146) &    (0.101) &   (0.0737) \\
		
		Experience &  -0.412*** &  -0.450*** &  -0.489*** &  -0.494*** \\
		
		&   (0.0337) &   (0.0871) &   (0.0247) &   (0.0344) \\
		
		Experience-Squared &   0.000848 &  -0.000681 & 0.00457*** &  0.00449** \\
		
		&  (0.00165) &  (0.00496) &  (0.00122) &  (0.00192) \\
		
		Black indicator &  -0.945*** &  -0.926*** &  -0.886*** &  -0.753*** \\
		
		&   (0.0939) &    (0.162) &    (0.113) &   (0.0701) \\
		
		Constant &   16.60*** &   16.42*** &   17.00*** &   16.68*** \\
		
		&    (0.242) &    (0.393) &    (0.173) &    (0.211) \\
		\hline
		& \multicolumn{ 4}{c}{{\bf Second-stage estimates}} \\
		\hline
		Education &   0.157*** &   0.176*** &   0.268*** &      0.104 \\
		
		&   (0.0524) &   (0.0521) &   (0.0271) &   (0.0662) \\
		
		Experience &   0.119*** &   0.120*** &   0.180*** &  0.0932*** \\
		
		&   (0.0227) &   (0.0248) &   (0.0140) &   (0.0341) \\
		
		Experience-Squared & -0.00236*** & -0.00201*** & -0.00337*** & -0.00221*** \\
		
		& (0.000347) & (0.000347) & (0.000352) & (0.000438) \\
		
		Black indicator &   -0.123** &   -0.110** &   -0.00925 &  -0.148*** \\
		
		&   (0.0520) &   (0.0519) &   (0.0342) &   (0.0469) \\
		
		Constant &   3.237*** &   2.698*** &   1.400*** &   4.360*** \\
		
		&    (0.883) &    (0.870) &    (0.466) &    (1.119) \\
		\hline
		Observations &      3,010 &      3,010 &      3,010 &      3,010 \\
		\hline
	\end{tabular}  
	
	Source: Card (1995). 
	Notes: Standard errors in parentheses. SE robust for OLS estimates. *** $p<0.01$, ** $p<0.05$, * $p<0.1$.  Regional and geographic dummies are used but omitted.						
	
\end{table}

Table \ref{table:card} shows the results of the first-stage to check if the IV are valid, together with the estimated second-stage results. The first column corresponds to the 2SLS mean model and the next ones are the regressions proposed for IVQR for $\tau\in\{0.25,0.50,0.75\}$. The results shows that the first instrument (lived near 2-year college in 1966) is not relevant for the low quantiles and the mean but it is significant for middle and high quantiles. Also, note that although the second instrument (lived near 4-year college in 1966) rejects the null hypothesis for the  mean, this variable has different degree of significance across quantiles. In particular, this is for $\tau=0.75$ where the instrument is relevant only at 10\% significance. These results are very important since although the proximity to the university seems to be a valid instrument to identify the causal effect of education on the mean, our test also indicates a certain limitation when the object of study is to evaluate the impact on the lower part of conditional distribution of wages. Therefore, this alerts for the quality of the asymptotic properties of the IVQR estimates in the presence of invalid instruments.

\section{Conclusions}\label{conclusion}

This paper proposes a first-stage model and a testing procedure to evaluate the degree of association between the IV and the endogenous regressor(s) in the IVQR estimator. The procedure developed here allows to evaluate instruments in a similar vein to that in 2SLS models for the conditional average, that is, by looking at the statistical significance of the instruments in the first-stage regression. In turn, this will allow to investigate IV validity for specific quantiles. Nevertheless, due to the requirement of consistent estimation of the weights in the first stage, it is important to notice that the testing requires the availability of at least of instrument. This caveat for QR IV models leads to two conclusions. First, it is difficult to derive a complete analogous F-statistic type rule-of-thumb for categorizing weak instruments. We leave this problem to future research. Second, we strongly suggest the use of weak identification robust inference procedures for QR models for practical work applying QR instrumental variables.
Monte Carlo experiments clearly illustrate that one may encounter cases where the IV are not valid for the mean, but are still valid for some quantiles. The same issue appears in the empirical application.

The analysis may be extended in the following two directions. First, this approach can be used to identify quantile-specific treatment effects, where an IV estimate being significant at some quantiles corresponds to a particular effect of a treatment. Second, the procedure outlined here could be combined with the second-stage inference to produce statistics similar to the \citet{Staiger97} F-statistics rule-of-thumb. In particular, to study weak instruments issues in QR models. 

\newpage

\appendix

\section*{Appendix A: Proofs}

\begin{proof}[Proof of Lemma \ref{lemmaJ}.]
	First, consider an estimator of the parameter $\mu$ using the true weighting matrix $V$ as
	\begin{equation}
	V= 
	\begin{bmatrix}
	f_{1} &  &  \\
	& \ddots &  \\
	&  & f_{n}
	\end{bmatrix},
	\end{equation}
that is given by the following
	\begin{equation*}
	\tilde{\mu}=(W' V W)^{-1}W' V D,
	\end{equation*}
	where $W=[X,Z]$. Replacing $D$ by $(W\mu_0 + \varepsilon)$ in the definition of $\tilde{\mu}$ we have that
	\begin{equation*}
	\sqrt{n}(\tilde{\mu}-\mu)=\left(\frac{W' V W}{n} \right)^{-1} \frac{W' V \varepsilon}{\sqrt{n}}.
	\end{equation*}
	
	By the Slutsky's Theorem, the proof of the lemma requires showing that
	\begin{equation}\label{eq:1}
	\frac{W' V W}{n} \stackrel{p}{\rightarrow}  \Omega_{f},
	\end{equation}
	and
	\begin{equation}\label{eq:2}
	\frac{W' V \varepsilon}{\sqrt{n}} \stackrel{d}{\rightarrow} N(0, \Omega_{f\sigma }).
	\end{equation}
	
	To show \eqref{eq:1}, its left side has the $(j, k)$ element given by
	\begin{equation*}
	\frac{1}{n} \sum_{i=1}^{n} f_{i} w_{ij}w_{ik} \stackrel{p}{\rightarrow} \E \left[ f_{i} w_{ij}w_{ik}  \right],
	\end{equation*}
	by the Law of Large Numbers and Assumption \ref{A2}. To show \eqref{eq:2}, first note that
	\begin{equation*}
	\E\left[ W' V \varepsilon \right] = \E\left[ W' V \E[\varepsilon | W] \right] =0,
	\end{equation*}
	by Assumption \ref{A2}. Furthermore, $W' V \varepsilon$ is a sum of i.i.d. random vectors $f_{\theta_0}(s_i) \cdot w_{i} \cdot \varepsilon_{i}$ with common covariance matrix having the $(j, k)$ element
	\begin{align*}
	Cov \left( f_{i} w_{ij} \varepsilon_{i}, f_{i} w_{ik} \varepsilon_{i}  \right) & = \E\left[ f_{i}^2 w_{ij}w_{ik} \varepsilon_{i}^{2} \right] = \E\left[ f_{i}^2 w_{ij}w_{ik} \E[\varepsilon_{i}^{2}|w_{i}] \right]\\
	& =  \E\left[ f_{i}^2 w_{ij}w_{ik} \sigma_{i}^{2} \right].
	\end{align*}
	Thus, each vector $f_{i} \cdot w_{i}\cdot \varepsilon_{i}$ has covariance matrix $\Omega_{f \sigma}$. Therefore, by the Multivariate Central Limit Theorem, \eqref{eq:2} holds.
	
	Finally, we have to show that using estimated weights does not affect the liming distribution. To establish that consider the estimator with the estimated weights as following
	\begin{equation*}
	\hat{\mu}=(W' \hat{V} W)^{-1}W' \hat{V} D,
	\end{equation*}
	such that
	\begin{equation}\label{eq:3}
	\sqrt{n}(\hat{\mu}-\mu)= \left(\frac{W' \hat{V} W}{n}  \right) \frac{W' \hat{V} \varepsilon}{\sqrt{n}} .
	\end{equation}
	
	First, we show that 
	\begin{equation}\label{eq:4}
	\frac{W' \hat{V} \varepsilon}{\sqrt{n}} - \frac{W'  V \varepsilon}{\sqrt{n}}\stackrel{p}{\rightarrow} 0.
	\end{equation}
	
	Note that 
	\begin{equation}\label{eq:app aux1}
	\frac{W' (\hat{V}-V) \varepsilon }{\sqrt{n}} = n^{-1/2} \sum_{i=1}^{n} w_{i} \varepsilon_{i} \left( \hat{f}_{i} - f_{i} \right).
	\end{equation}
	
	We want to show that the right hand side of \eqref{eq:app aux1} is $o_{p}(1)$.
	Using the sparsity function estimator in \eqref{eq:sparsity} along with some calculations, we have
	that
	\begin{equation*}
	\hat{f}_{i} = f_{i} + \frac{2h_{n}}{f_{i}^{2}} s_{i} (\hat{\theta}-\theta) + o_{p}((nh^{2})^{-2/3}).
	\end{equation*}
	We refer the reader to \citet{OtaKatoHara19} for details on the remainder term. 
	
	Hence, using the previous equation, the $j$th component of the right hand side of equation \eqref{eq:app aux1} can be written as
	\begin{equation*}
	\sqrt{n}(\hat{\theta}_{j} -\theta_{0,j}) 2h_{n} \frac{1}{n}  \sum_{i=1}^{n} \frac{1}{f_{i}^{2}}w_{ij} \varepsilon_{i}.
	\end{equation*}
	
	The first factor $\sqrt{n}(\hat{\theta}_{j} -\theta_{0,j})=o_{p}(1)$ by Assumption \ref{A1} and CH. Moreover, note that the average of the i.i.d. variables $f_{i}^{-2}w_{i} \varepsilon_{i}$ obeys the Law of Large Numbers by the moment restrictions in Assumption \ref{A2}, and the result follows.
	
	
	Next, we show that
	\begin{equation}\label{eq:5}
	\frac{W' \hat{V} W}{n} - \frac{W'  V W}{ n}\stackrel{p}{\rightarrow} 0,
	\end{equation}
	which follows from the same argument as above.
	
	The convergences \eqref{eq:4} and \eqref{eq:5} are enough to show that the right-hand side of \eqref{eq:3} satisfies
	\begin{equation*}
	\left(\frac{W' \hat{V} W}{n}  \right) \frac{W' \hat{V} \varepsilon}{\sqrt{n}} - \left(\frac{W' V W}{n}  \right) \frac{W' V \varepsilon}{\sqrt{n}} \stackrel{p}{\rightarrow} 0
	\end{equation*}
	just by making simple use of the equality
	\begin{equation*}
	\hat{a} \hat{b} - ab = \hat{a}(\hat{b} - b) + (\hat{a}- a)b.
	\end{equation*}
	Finally, Slutsky's theorem yields the result.
\end{proof}


\begin{proof}[Proof of Proposition \ref{prop:delta}.]
The proof of this result is simple. It follows from observing that  by Lemma \ref{lemmaJ},
\begin{equation*}
\sqrt{n}(\hat{\mu}-\mu_0) \stackrel{d}{\rightarrow} N\left(\bm 0,V(\mu_0)\right).
\end{equation*}
Notice that $R \mu= \delta_{1}$, hence under the null hypothesis,
\begin{equation*}
\sqrt{n}(R\hat{\mu}- \bm 0) \stackrel{d}{\rightarrow} N\left(\bm 0, R V(\mu_{0}) R'\right).
\end{equation*}
Let $\hat{V}(\hat\mu)$ be a consistent estimator of $V(\mu_0)$, and $V_{\delta_{1}}:=R V(\mu_0) R' $, then by the Slutsky's theorem,
\begin{equation*}
T_n=n \left( \hat\delta_1 \right)' \{ V_{\delta_{1}} \}^{-1} \left( \hat\delta_1 \right)	\stackrel{d}{\rightarrow} \chi^2_{p_1}(\bm{a}_{p_1}).
\end{equation*}

\end{proof}

\vspace{1cm}
\section*{Appendix B: Examples of weighting factors}

\subsubsection*{1. Location model}

Consider a pure location model, using two equations
\begin{align*}
	y & = d + u,\\
	d & = a z + v,
\end{align*}
with $(u,v)\sim N(0,0,1,1,\rho)$ a bivariate normal with zero mean, unit variance and correlation parameter $\rho$ and $z\sim N(0,1)$. Then, it follows that $d\sim N(0,1+a^2)$ and $y\sim N(0,2+a^2+2\rho)$. 

Consider now the model where we condition on both $(d,z)$. For this case, $u|d,z\sim N(\rho v, (1-\rho^2))$ by the marginal of the bivariate normal density. Then, 
$$Q_\tau(u|d,z)=\rho v+\sqrt{1-\rho^2}\Phi^{-1}(\tau).$$

Then, $u_\tau=y-	Q_\tau(y|d,z)=u-Q_\tau(u|d,z)$. Note that $\E(u_\tau|d,z)=\E(u_\tau|d,z)-Q_\tau(u|d,z)=-\sqrt{1-\rho^2}\Phi^{-1}(\tau)$. Thus, the density is
$$f_{u_\tau}(U|d,z)=\frac{1}{\sqrt{1-\rho^2}}\phi\left(\frac{U+\sqrt{1-\rho^2}\Phi^{-1}(\tau)}{\sqrt{1-\rho^2}}\right),$$
where $\phi()$ is the density function of a standard normal. If we evaluate it at $0$,
$$f_{u_\tau}(0|d,z)=\frac{1}{\sqrt{1-\rho^2}}\phi\left(\Phi^{-1}(\tau)\right).$$

Now consider the joint density of $(u,d)\sim N(0,0,1,1+a^2,\kappa)$, where $\kappa =\frac{\rho}{\sqrt{1+a^2}}$. Then, it follows that $u|d\sim N(\E(u|d),Var(u|d))$, where $\E(u|d)=\kappa d$ and $Var(u|d)=(1-\kappa^2)$.

As such, we can obtain the quantiles of interest,
$$Q_\tau(y|d)=d+\kappa d+\Phi^{-1}(\tau)(1-\kappa^2)^{1/2}.$$

Note that without endogeneity, i.e. $\rho=0$, then $\kappa=0$, and the correct $\tau$-quantile model should be 	
$$Q_\tau(y|d,\rho=0)=d+\Phi^{-1}(\tau).$$

Now, $u^*_\tau=y-Q_\tau(y|d)=d+u-(d+\kappa d+\Phi^{-1}(\tau)(1-\kappa^2)^{1/2})=u-\kappa d-\Phi^{-1}(\tau)(1-\kappa^2)^{1/2}$.  Then, $\E(u^*_\tau|d)=-\Phi^{-1}(\tau)(1-\kappa^2)^{1/2}$, and $Var(u^*_\tau|d)=Var(u|d)=(1-\kappa^2)$.

Then,
$$f_{u^*_\tau}(U|d)=\frac{1}{\sqrt{(1-\kappa^2)}}\phi\left(\frac{U-\E(u^*_\tau|d)}{\sqrt{Var(u^*_\tau|d)}}\right),$$
such that,
$$f_{u^*_\tau}(0|d)=\frac{1}{\sqrt{(1-\kappa^2)}}	
\phi\left(\Phi^{-1}(\tau)\right).$$

In all cases, $f_{u^*_\tau}(0|d)$ and $f_{u_\tau}(0|d,z)$ are constant that do not change with $d$ or $z$.
It is interesting to evaluate when $a=0$, such that $(1-\kappa^2)=(1-\rho^2)$. Note that in this case, $f_{u^*_\tau}(0|d)=f_{u_\tau}(0|d,z)$.

\subsubsection*{2. Location-scale model 1}

Now consider a location-scale model of the form
\begin{align*}
	y & = d + (1+cd)u,\\
	d & = az + v,
\end{align*}
where $a$ and $c$ are parameters. As in the previous case $(u,v)\sim N(0,0,1,1,\rho)$.  Then, $u|(d,z)\sim u|v\sim N(\rho v,1-\rho^2)$. Thus, $Q_\tau(u|d,z)=\rho v+\sqrt{1-\rho^2}\Phi^{-1}(\tau)$. Note that it does not depend on $z$.

In this case, $Q_\tau(y|d,z)=d+(1+cd)Q_\tau(u|d,z)$, and then, $u_\tau=y-Q_\tau(y|d,z)=(1+cd)(u-Q_\tau(u|d,z))$.

As such, we can obtain,
$$f_{u_\tau}(U|d,z)=\frac{1}{|1+cd|\sqrt{1-\rho^2}}\phi\left(\frac{U+(1+cd)\sqrt{1-\rho^2}\Phi^{-1}(\tau)}{(1+cd)\sqrt{1-\rho^2}}\right).$$
If we evaluate it at $0$,
$$f_{u_\tau}(0|d,z)=\frac{1}{|1+cd|\sqrt{1-\rho^2}}\phi\left(\Phi^{-1}(\tau)\right).$$
Note that this depends $d$, and then, the weights are not uniform.

Now, consider the of $u|d$. Consider first the joint distribution of $(u,d)\sim N(0,0,1,1+a^2,\kappa)$ where $\kappa=\rho/\sqrt{1+a^2}$. Now, $u|d\sim N(\kappa d, (1-\kappa^2))$, then $\E(u|d)=\kappa d$ and
$Var(u|d)=(1-\kappa^2)$. 

For this case let	
$u^*_\tau=y-Q_\tau(y|d)=d+(1+cd)u-d-(1+cd)Q_\tau(u|d)=(1+cd)(u-Q_\tau(u|d))$. Since $u|d$ is Gaussian then	$(1+cd)(u-\kappa d-\Phi^{-1}(\tau)(1-\kappa^2)^{1/2})$. Then, $\E(u^*_\tau|d)=(1+cd)(-\Phi^{-1}(\tau)(1-\kappa^2)^{1/2})$ and 
$Var(u^*_\tau|d)=(1+cd)^2(1-\kappa^2)$.	 As such, we can obtain,

$$f_{u^*_\tau}(U|d)=\frac{1}{|1+cd|\sqrt{1-\kappa^2}}\phi\left(\frac{U+(1+cd)(1-\kappa^2)^{1/2}\Phi^{-1}(\tau)}{(1+cd)(1-\kappa^2)^{1/2}}\right).$$
If we evaluate it at $0$,
$$f_{u^*_\tau}(0|d)=\frac{1}{|1+cd|\sqrt{1-\kappa^2}}\phi\left(\Phi^{-1}(\tau)\right).$$

Note that both $f_{u_\tau}(0|d,z)$ and $f_{u^*_\tau}(0|d)$ share the same relationship with $d$. In fact, the weighting procedure will be equivalent, as they are proportional to each other.

\subsubsection*{3. Location-scale model 2}

Now consider a location-scale model where both the first and second stage are affected in the variance component,
\begin{align*}
	y &= d + (1+cd)u,\\
	d &= az + (1+bz)v,
\end{align*}
where $a$, $b$, and $c$ are parameters. As in the previous case $(u,v)\sim N(0,0,1,1,\rho)$. Define $w=(1+bz)v$ and note that $(u,w|z)\sim N(0,0,1,(1+bz)^2,\rho)$. Then, $u|d,z\sim u|w,z\sim N(\rho v,1-\rho^2)$. Thus, $Q_\tau(u|d,z)=\rho v+\sqrt{1-\rho^2}\Phi^{-1}(\tau)$. Note that it does not depend on $b$.

In this case, $Q_\tau(y|d,z)=d+(1+cd)Q_\tau(u|d,z)$, and then, $u_\tau=y-Q_\tau(y|d,z)=(1+cd)(u-Q_\tau(u|d,z))$.

As such, we can obtain,
$$f_{u_\tau}(U|d,z)=\frac{1}{|1+cd|\sqrt{1-\rho^2}}\phi\left(\frac{U+(1+cd)\sqrt{1-\rho^2}\Phi^{-1}(\tau)}{(1+cd)\sqrt{1-\rho^2}}\right).$$
If we evaluate it at $0$,
$$f_{u_\tau}(0|d,z)=\frac{1}{|1+cd|\sqrt{1-\rho^2}}\phi\left(\Phi^{-1}(\tau)\right).$$
Note that this depends $d$, and then, the weights are not uniform.

Now, it is not standard to obtain the distribution of $u|d$. To exemplify this, suppose $z=\left\lbrace0,1\right\rbrace$ is a simple binary variable with $p=Pr(z=1)$ and independent of $(u,v)$. Then, the joint density is $f(u,v,z)=\phi_\rho(u,v)p^{z}(1-p)^{1-z}$ and using the Jacobian transformation we obtain:
$$f(u,d,z) =  \frac{1}{|1+bz|} \phi_\rho\left(u,\frac{d-az}{1+bz}\right)p^{z}(1-p)^{1-z}.  		
$$

Therefore,
$$f(u,d) =  \phi_\rho\left(u,d\right)(1-p) +\frac{1}{|1+b|} \phi_\rho\left(u,\frac{d-a}{1+b}\right)p,	
$$
and
$$f(d) =  \phi(d)(1-p) + \frac{1}{|1+b|}\phi\left(\frac{d-a}{1+b}\right)p.$$

Putting all that together, the conditional density is
$$f(u|d) =  \frac{\phi_\rho\left(u,d\right)(1-p) +\frac{1}{|1+b|} \phi_\rho\left(u,\frac{d-a}{1+b}\right)p}{\phi(d)(1-p) + \frac{1}{|1+b|}\phi\left(\frac{d-a}{1+b}\right)p}.	
$$

If we assume that $p=\frac{|1+b|}{1+|1+b|}$ this expression simplifies to
$$f(u|d) =  \frac{\phi_\rho\left(u,d\right) + \phi_\rho\left(u,\frac{d-a}{1+b}\right)}					  {\phi(d) + \phi\left(\frac{d-a}{1+b}\right)}.	
$$

We can rewrite this as a function of standard normal densities noting that $\phi_\rho(u,d) = \phi_\rho(u|d)\phi(d)$ with $\phi_\rho(u|d)=\frac{1}{\sqrt{1-\rho^2}}\phi\left(\frac{u-\rho d}{\sqrt{1-\rho^2}}\right)$, then
$$f(u|d) =  \frac{1}{\sqrt{1-\rho^2}}\phi\left(\frac{u-\rho d}{\sqrt{1-\rho^2}}\right)\omega(d) + \frac{1}{\sqrt{1-\rho^2}}\phi\left(\frac{u-\rho \frac{d-a}{1+b}}{\sqrt{1-\rho^2}}\right)(1-\omega(d)),
$$
where $\omega(d)= \frac{\phi(d)}{\phi(d) + \phi\left(\frac{d-a}{1+b}\right)}$. Therefore, conditional on $d$ this density is a Gaussian mixture of two distributions with different means. Two particular cases are: (i) $\rho=0$ (exogeneity) where $f(u|d) = \phi(u)$; (ii) $a=b=0$ ($d$ and $z$ unrelated) which reduces to $f(u|d) = \phi_\rho(u|d)$. Obviously, in the rest of the cases $Q_\tau(u|d)$ does not have an explicit analytical solution and therefore neither $u^*_\tau=y-Q_\tau(y|d)=(1+cd)(u-Q_\tau(u|d))$.

The interesting feature to notice is that in all cases, the distribution of $u^*_\tau$ depends basically on $d$, and $(1+cd)$ should be used to standardize its density function in a similar way to $u_\tau$.

\end{document}